\documentstyle[12pt]{article}
 1
\title{\marginpar{\vspace{-1in}\hspace{-1in}\small KFT U{\L} 8/95}
The H-Boltzmann's Theorem\protect\\
for Magnetic Systems}
\author{S. Malinowski\vspace{2ex}\\
University of {\L}\'od\'z\\
Departament of Theoretical Physics\\
ul.~Pomorska 149/153\\
90-236 {\L}\'od\'z, Poland }
\date{June 1995}
\begin{document}
\maketitle
\thispagestyle{empty}
\vfill
\begin{abstract}
In  a  time-reversal   non-invariant   system   the   entropy
production, dS/dt, may be negative. Close to  the critical  point
of medium it is positive: larger or smaller as compared with  one
above critical point depending on whether  the  transitions  from
lower states to higher states  of  the  system  are  more  (less)
probable than the reverse transitions.
\end{abstract}
\vfill
\newpage
\setcounter{page}{2}

The entropy S is defined [1] as follows:
\begin{equation}
S=-k\sum^{\infty}_{n=1}P_n\ln P_n
\end{equation}
where k is Boltzmann's constant and
\begin{equation}
P_n=\frac{\exp(-E_n/kT)}{\sum^{\infty}_{m=1}\exp(-E_m/kT)}
\end{equation}
is the Gibbs distribution of the energy levels $E_n$. To clarify the
considerations we will use  the  following  arrangement  for  the
energy levels:
\begin{equation}
E_1\le E_2\le\ldots\le E_{\infty}.
\end{equation}
Hence, for the probability distributions we obtain the
relations:
\begin{equation}
1\ge P_1\ge P_2\ge\ldots\ge P_{\infty}\ge0.
\end{equation}
The time derivative of entropy is equal to
\begin{equation}
\frac{{\rm d}S}{{\rm d}t}=-k\sum^{\infty}_{n=1}\frac{{\rm
d}P_n}{{\rm d}t}\ln P_n
\end{equation}
because
\begin{displaymath}
\frac{\rm d}{{\rm d}t}\sum^{\infty}_{n=1}P_n=0.
\end{displaymath}
{}From Boltzmann's equation we have
\begin{equation}
\frac{{\rm d}P_n}{{\rm
d}t}=\sum^{\infty}_{m=1}(W_{m,n}P_m-W_{n,m}P_n),
\end{equation}
where $W_{m,n}$ is the transition probability from the
state $\psi_m$ (level
$E_m$) to the state $\psi_n$  (level $E_n$).
Hence, for ${\rm d}S/{\rm d}t$ we have the following
equivalent formulae:
\setcounter{equation}{7}
$$
\frac{{\rm d}S}{{\rm
d}t}=-k\sum^{\infty}_{n=1}\sum^{\infty}_{m=1}(W_{m,n}P_m\ln
P_n-W_{n,m}P_n\ln P_n)
\eqno(\theequation a)
$$
$$
=k\sum^{\infty}_{n=1}\sum^{\infty}_{m=1}W_{n,m}P_n\ln
\frac{P_n}{P_m}
\eqno(\theequation b)
$$
$$
=k\sum^{\infty}_{n=1}\sum^{\infty}_{i=n+1}(W_{n,n+i}P_n-W_{n+i,n}P_{n+i})
\ln\frac{P_n}{P_{n+i}}
\eqno(\theequation c)
$$
Let us compare the  entropy  production  in  the  same  sample
before and after breaking its time-inversion  symmetry.  Consider
the electronic system in the temperature $T_1$ above, $T_c
< T_1$, and
$T_2$  below, $T_2 < T_c$ , a critical temperature $T_c$
is  an  example  of
such medium.

In temperature $T_1$  the  matrix of transition probabilities is a
symmetric one:
\begin{equation}
W_{n,n+i}(0)=W_{n+i,n}(0)\equiv W^{(s)}_{n,n+i},
\end{equation}
while in temperature $T_2$  it is not symmetric  and  moreover  it is
some function of spontaneous magnetization $\bf M$ of the
sample:
\begin{equation}
W_{n,n+i}({\bf M})=W^{(s)}_{n,n+i}({\bf
M})+W^{(a)}_{n,n+i}({\bf M}).
\end{equation}
Hence, we have
\begin{equation}
W_{n+i,n}({\bf M})=W^{(s)}_{n,n+i}({\bf
M})-W^{(a)}_{n,n+i}({\bf M}).
\end{equation}
The time-inversion operator transforms $W_{n,n+i}({\bf M})$   into
$W_{n+i,n}(-{\bf M})$. It means that the $W^{(s)}$
is  an even function of $\bf M$,
\begin{equation}
W^{(s)}_{n,n+i}({\bf M})=W^{(s)}_{n,n+i}(-{\bf M})
\end{equation}
and   $W^{(a)}$    - an odd one,
\begin{equation}
W^{(a)}_{n,n+i}({\bf M})=-W^{(a)}_{n,n+i}(-{\bf M}).
\end{equation}
The other properties are  following:
\begin{equation}
\lim_{{\bf M}\rightarrow0}W^{(s)}_{n,n+i}({\bf
M})\rightarrow W^{(s)}_{n,n+i}(0),
\end{equation}
\begin{equation}
\lim_{{\bf M}\rightarrow0}W^{(a)}_{n,n+i}({\bf M})=0,
\end{equation}
\begin{equation}
|W^{(a)}_{n,n+i}({\bf M})|\le W^{(s)}_{n,n+i}({\bf M}).
\end{equation}
The latter follows from the definition of the $W_{n,n+i}$.

The $W^{(a)}_{n,n+i}({\bf M})$ may be positive or negative.
If it is positive:
\begin{equation}
W^{(a)}_{n,n+i}({\bf M})>0
\end{equation}
then the probability of transition from the lower state $\psi_n$  to the
higher state $\psi_{n+i}$ is larger than the probability  of  the  reverse
transition. If it is negative:
\begin{equation}
W^{(a)}_{n,n+i}({\bf M})<0
\end{equation}

then we  are  dealing  with  the  reverse  relation  betweem  the
probabilities of transition.

Inserting the expressions (9) and (10) to the formula (7c) we
obtain
\begin{displaymath}
\frac{{\rm d}S({\bf M})}{{\rm
d}t}=k\sum^{\infty}_{n=1}\sum^{\infty}_{i=n+1}
[W^{(s)}_{n,n+1}({\bf M})(P_n-P_{n+i})+W^{(a)}_{n,n+i}({\bf
M})(P_n+P_{n+i})]\ln\frac{P_n}{P_{n+i}}
\end{displaymath}
\begin{equation}
=\frac{{\rm d}S({\bf M})_{\rm inv}}{{\rm
d}t}+\frac{{\rm d}S({\bf M})_{\rm noninv}}{{\rm
d}t}
\end{equation}
where the invariant part of ${\rm d}S({\bf M})/{\rm d}t$
with  respect  to  the  time
inversion is always positive:
\begin{equation}
\frac{{\rm d}S({\bf M})_{\rm inv}}{{\rm
d}t}=k\sum^{\infty}_{n=1}
\sum^{\infty}_{i=n+1}W^{(s)}_{n,n+i}({\bf M})(P_n-P_{n+i})
\ln\frac{P_n}{P_{n+i}}>0,
\end{equation}
and the noninvariant part
\begin{displaymath}
\frac{{\rm d}S({\bf M})_{\rm noninv}}{{\rm
d}t}=k\sum^{\infty}_{n=1}
\sum^{\infty}_{i=n+1}W^{(a)}_{n,n+i}({\bf M})(P_n+P_{n+i})
\ln\frac{P_n}{P_{n+i}}
\end{displaymath}
may acquire, by ineq. (16)  and  (17)  an arbitrary sign.\\[0.7cm]

IN CONCLUSION: Far apart from the critical  point  of  a  sample,
when  $W^{(a)}$  may achieve the absolute values  comparable  with  the
values of $W^{(s)}$ and almost all $W^{(a)}_{n,n+i}({\bf M})$
are  negative  the  entropy
production may be negative
\begin{equation}
\frac{{\rm d}S({\bf M})}{{\rm d}t}<0.
\end{equation}
Close to the critical point, due to the properties (13) and (14),
the entropy production is positive and for some samples is larger
than the one above critical point,
\begin{equation}
\frac{{\rm d}S({\bf M})}{{\rm d}t}>\frac{{\rm d}S(0)}{{\rm
d}t}>0
\end{equation}
and for the others is smaller
\begin{equation}
\frac{{\rm d}S(0)}{{\rm d}t}>\frac{{\rm d}S({\bf M})}{{\rm
d}t}>0.
\end{equation}

   Transport phenomena are examples  of  irreversible  processes.
The formulae (21) and (22) translated to  the  behaviour  of  the
electric resistivity $\rho$ in the  neighbourhood  of  critical  point
lead to the inequality
\begin{equation}
\rho({\bf M})<\rho(0)
\end{equation}
and
\begin{equation}
\rho(0)>\rho({\bf M}),
\end{equation}

respectively, which is well-known to experimentators [2] (see the
papers listed in [2]). In this situation the above generalization
of  Boltzmann's  theorem  can  be  treated  as   the   additional
interpretation of phase transitions. The explicite  form  of  the
transition probability $W_{n,m}$  satisfying the requirement  (9)  will
 be subject of the next paper.\\[0.7cm]

 Acknowledgement\\[0.3cm]
    I wish to thank Professor P. Kosiñski for fruitful discussions
 and reading the manuscript \\[0.5cm]

\end{document}